\documentclass[aps,prl,twocolumn,showpacs,superscriptaddress,groupedaddress]{revtex4-1}
\usepackage{graphicx}
\usepackage{amssymb}
\usepackage{amsmath}
\usepackage{verbatim}
\usepackage{ulem}
\usepackage{bm}
\usepackage{mathtools}
\usepackage{hyperref}
\hypersetup{ pdfborder = {0 0 0}}
\DeclareMathAlphabet\mathbfcal{OMS}{cmsy}{b}{n}

\begin{document}
\title{Dynamical control of atoms with polarized bichromatic weak field}
\author{A. Kani}
\author{Harshawardhan Wanare}
\affiliation{Department of Physics, Indian Institute of Technology, Kanpur 208016, India}

\begin{abstract}
We propose ultranarrow dynamical control of population oscillation (PO) between ground states through the polarization content of an input bichromatic field. Appropriate engineering of classical interference between optical fields results in PO arising exclusively from optical pumping. Contrary to the expected broad spectral response associated with optical pumping, we obtain subnatural linewidth in complete absence of quantum interference. The ellipticity of the light polarizations can be used for temporal shaping of the PO leading to generation of multiple sidebands even at low light level.
\end{abstract}
\maketitle
All subnatural response involving atom-light interaction is based on quantum interference (QI) effects~\cite{ultranarrow1,EIT}. In the most elementary configuration QI involves at least two transitions sharing a common state, for example, the three-level $\Lambda$-system, wherein, QI traps the atom in a coherent superposition of the ground states making the medium transparent at the two-photon Raman condition. The subnatural response arises from the associated coherence generated between the long-lived ground states~\cite{gsc1,gsc2}, and limited by the ground state decoherence. The large magnitude of the ground state coherence does not affect the atomic population in these states, and the relative strength of the fields coupling the two arms of the $\Lambda$-system determines the population distribution. QI effects by nature are fragile and hence decohere quickly, whereas, population transfer is accompanied with robust experimental signatures. The coherent population oscillation (CPO) phenomena is widely observed in two-level atoms involving bichromatic excitation~\cite{CPO1,CPO2,CPO3,CPO4}, wherein, one field is strong and saturates the transition allowing the other weak field to modulate only a fraction of the population between the ground and excited states leading to limited optical depth within a narrow spectral range. The origin of CPO in multi-level systems~\cite{cpolambda1,cpolambda2,cpolambda3,cpolambda4} continues to be intrigue us with regard to the role of QI in such systems. The narrowband nature of CPO could either be associated with the underlying QI or with the significant difference in the Rabi frequency associated with the weak probe field in comparison to the strong saturating field.

\begin{figure}[ht]
\includegraphics[width=\linewidth]{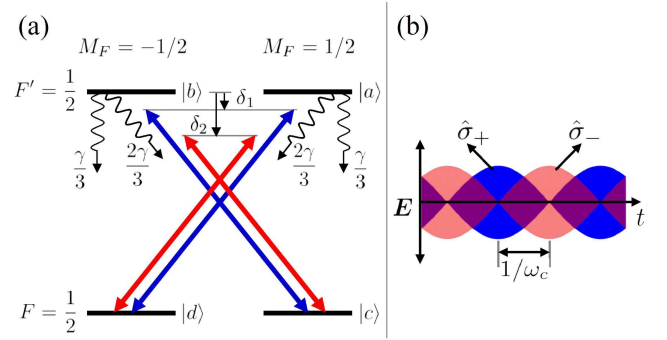}
\caption{(a) $1/2\to1/2$-system interacting with bichromatic optical field. (b) The amplitude of $\hat{\sigma}_\pm$ components of orthogonal linearly polarized bichromatic field oscillating at $\omega_c/2=(\omega_1-\omega_2)/2$ with $\pi/2$ phase difference.}\label{OPint}
\end{figure}

In contrast to the above, we demonstrate \textit{classical interference} in atom-light interaction leading to \textit{complete} periodic population transfer between ground states in absence of any QI and without taking recourse to strong saturating fields. This leads to immense optical depth within subnatural linewidth involving nearly complete population transfer, in contrast to the quantum superposition based effects wherein a plethora of external influences quite severely limit QI and hence the resulting performance. The population oscillation (PO) also leads to nonlinear frequency generation, and the temporal evolution of the population governs the generated frequencies which can be controlled by changing the amplitudes as well as the polarizations content of the bichromatic field. The PO here arises purely from optical pumping (OP)~\cite{opticalp1,opticalp2}, and we show that the OP rate dictates the subnatural width of PO.

We consider a simple four level atomic system $F=1/2\to F'= 1/2$, interacting with a weak bichromatic light field of frequencies $\omega_1$ and $\omega_2$. The circularly polarized components $(\hat{\sigma}_\pm)$ associated with the light fields couple the atomic states in accordance to the electric dipole selection rule, as shown in Fig~\ref{OPint}(a). This results in effectively two independent two-level systems coupled only through spontaneous emission, which can not create any coherence between the two, and hence, QI does not play any role in this configuration. The polarization content of the two fields involve two orthogonal linearly polarized fields ${\bm E}_1=E_0\hat{x}\cos(k_1z-\omega_1t)$ and ${\bm E}_2=E_0\hat{y}\cos(k_2z-\omega_2t+\phi)$, where, $E_0$ is the amplitude of the fields and $\phi$ is the initial phase difference between the two input fields propagating along $\hat{z}$ direction. For obtaining maximum PO, the amplitude of the two fields are taken to be equal. The electric fields in the atomic spherical basis can be written as 
\begin{equation}
\begin{aligned}
{\bm E}_1&=\left(-\hat{\sigma}_++\hat{\sigma}_-\right)\frac{E_0}{2\sqrt{2}}e^{i(k_1z-\omega_1t)}+c.c,\\
{\bm E}_2&=\left(-i\hat{\sigma}_+-i\hat{\sigma}_-\right)\frac{E_0}{2\sqrt{2}}e^{i(k_2z-\omega_2t+\phi)}+c.c.
\end{aligned}
\end{equation}
As the polarization of the two fields are chosen to be orthogonal, the two circularly polarized components oscillating at half the beat frequency $\omega_1-\omega_2$ have a phase difference of $\pi/2$ as shown in Fig.~\ref{OPint}(b). This phase difference between the two amplitude modulated $\hat{\sigma}_{+}$ and $\hat{\sigma}_-$ components is paramount to the resulting PO between the ground states even for weak excitation. Moreover, as OP plays a central role in these considerations, the beat frequency should be smaller than the rate of OP. This would result in the ground state population adiabatically following the amplitude variation of the two fields. If the two optical frequencies are well separated then OP has little influence on the two fields, and the induced dipole moment for weak excitation under this condition can be written as
\begin{equation}\label{paway}
\begin{aligned}
{\bm p}\propto\frac{iE_0e^{-i\omega_1 t}}{3\left(\gamma-2i\delta_1\right)}\hat{x}+\frac{iE_0 e^{-i\omega_2 t+i\phi}}{3\left(\gamma-2i\delta_2\right)}\hat{y}+c.c.
\end{aligned}
\end{equation}
where, $\gamma$ is the total spontaneous emission rate, $\delta_1$ and $\delta_2$ are the single photon detuning of the two fields. The above Eqn.~\eqref{paway} is obtained from the conventional perturbative analysis retaining only the terms linear in the magnitude $E_0$. The Eqn.~\eqref{paway} is not valid when the frequencies are close, ie., $\omega_1\approx\omega_2$.  The Master equation governing the the atomic system is
\begin{equation}\label{master}
\begin{aligned}
\dot{\rho}_{aa}&=-\gamma\rho_{aa}-i\Omega\left(\eta_+^*(t)\rho_{ad}-\eta_+(t)\rho_{da}\right),\\
\dot{\rho}_{bb}&=-\gamma\rho_{bb}+i\Omega\left(\eta_-^*(t)\rho_{bc}-\eta_-(t)\rho_{cb}\right),\\
\dot{\rho}_{cc}&=\frac{\gamma}{3}(\rho_{aa}+2\rho_{bb})-i\Omega\left(\eta_-^*(t)\rho_{bc}-\eta_-(t)\rho_{cb}\right),\\
\dot{\rho}_{dd}&=\frac{\gamma}{3}(2\rho_{aa}+\rho_{bb})+i\Omega\left(\eta_+^*(t)\rho_{ad}-\eta_+(t)\rho_{da}\right),\\
\dot{\rho}_{ad}&=-\left(\frac{\gamma}{2}+i\omega_0\right)\rho_{ad}+i\Omega\eta_+(t)(\rho_{dd}-\rho_{aa}),\\
\dot{\rho}_{bc}&=-\left(\frac{\gamma}{2}+i\omega_0\right)\rho_{bc}-i\Omega\eta_-(t)(\rho_{cc}-\rho_{bb}),
\end{aligned}
\end{equation}
where, $\omega_0$ is the atomic resonance frequency, $\Omega=|d_{eg}|E_0/(2\sqrt{2})$ is the Rabi frequency, $d_{eg}$ is the dipole moment between the ground and excited states, and $\eta_\pm(t)=\mp e^{-i\omega_1 t}-ie^{-i(\omega_2 t-\phi)}$ contains the explicit time dependence. In order to systematically capture the atomic response, we use the closed-loop decomposition introduced earlier in Ref~\cite{kaniEPL}. Both the two-level systems consist of closed-loop transition such as absorption of a $\omega_2$ photon followed by emission of $\omega_1$ photon ${\bf E}_2\uparrow\downarrow{\bf E}_1^*$ (or vice versa ${\bf E}_1\uparrow\downarrow{\bf E}_2^*$), with the elementary closed-loop frequency being $\omega_c=\omega_1-\omega_2$ accompanied by a closed-loop phase $\phi_c=\phi$. The closed-loop decomposition of the diagonal and the nonzero off-diagonal density matrix elements is given as
\begin{figure}[ht]
\includegraphics[clip,trim=.2cm .2cm .2cm .2cm,width=\linewidth]{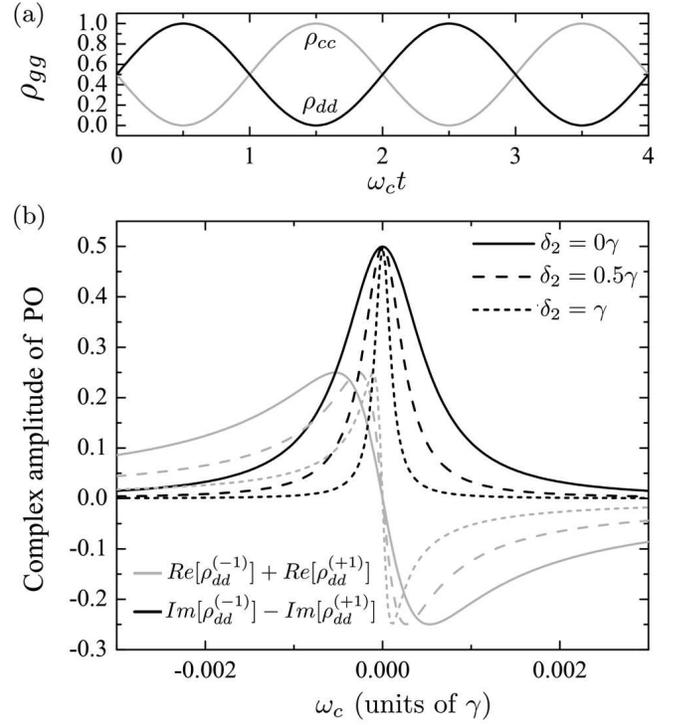}
\caption{(a) Ground states PO when $\omega_c\approx0$. (b) The complex amplitude of PO as a function of $\omega_c$ for weak excitation $\Omega=0.01\gamma$ at various single photon detuning $\delta_2$. Gray and black curves represent the real amplitude of in-phase and in quadrature components.}\label{rhocc2}
\end{figure}
\begin{equation}\label{WM1}
\begin{aligned}
\rho_{ii}&=\sum _{n=-\infty}^{+\infty}\rho_{ii}^{(n)}e^{in(\omega_c t-\phi_c)},\\
\rho_{eg}&=\sum _{n=-\infty}^{+\infty}\rho_{eg}^{(n)}e^{-i\omega_{1}t}e^{in(\omega_c t-\phi_c)},\\
\rho_{ge}&=\sum _{n=-\infty}^{+\infty}\rho_{ge}^{(n)}e^{i\omega_{1}t}e^{in(\omega_c t-\phi_c)}.
\end{aligned}
\end{equation}
\begin{figure}[ht]
\includegraphics[clip,trim=.2cm .2cm .2cm .2cm,width=\linewidth]{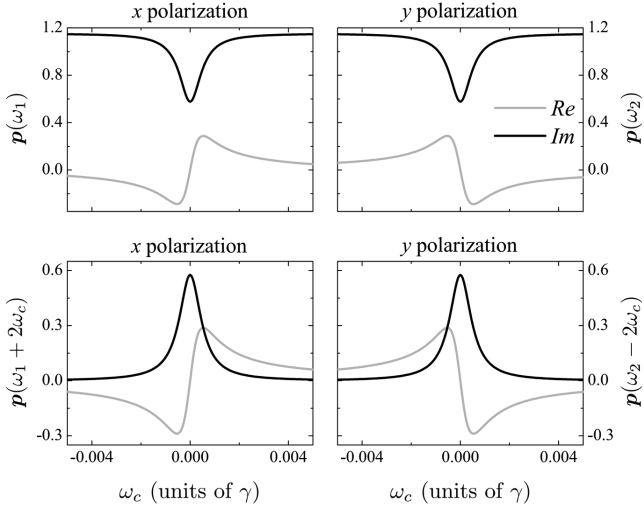}
\caption{Complex amplitude of the induced dipole moments at various harmonics as a function of $\omega_c$ for $\Omega=.01\gamma$ and $\delta_2=0$.}\label{WMfig}
\end{figure}
The above decomposition is valid for all strengths of the input fields, and leads to an infinite set of coupled equations, which can be solved numerically with the condition $\rho_{ij}^{(n\to\pm\infty)}=0$. The numerical solution for the ground state population $\rho_{gg}$ is shown in Fig.~\ref{rhocc2}(a) when the system adiabatically follows the amplitude variation of the $\hat{\sigma}_{\pm}$ components of the electric field shown in Fig.~\ref{OPint}(b) for $\omega_c\approx0$. For weak excitation $\Omega\ll\gamma$, $\rho_{ee}\approx0$, and the PO is between the ground states $\rho_{cc}\rightleftharpoons \rho_{dd}$, and only the dynamical terms $\rho_{gg}^{(\pm1)}e^{\pm i\omega_c t}$ contribute, which results in sinusoidal PO. The amplitude and phase of PO is governed by the complex amplitude $\rho_{gg}^{(\pm1)}$ and are shown in Fig.~\ref{rhocc2}(b) for various single photon detuning $\delta_2$. The subnatural linewidth associated with the PO can be enhanced by increasing the single photon detuning. The ground state PO is invariably accompanied with simultaneous nonlinear frequency generation at frequencies $\omega_i\pm\omega_c$. The complex amplitudes of the induced polarization at all the four frequencies are shown in Fig.~\ref{WMfig}. The polarization content of the generated frequencies are governed by the induced dipole moment vector, and the dipole moment at the frequency $\omega_1+2\omega_c$ has the polarization content corresponding to the field $\omega_1$, whereas, the dipole moment at the frequency $\omega_2-2\omega_c$ has the polarization corresponding to $\omega_2$.

We present the analytical solution of this nonlinear system under weak excitation. The numerical results are used to identify the most significant terms $\rho_{ij}^{(n)}$ relevant under weak excitation, and thus we truncate the closed-loop expansion retaining only the following terms $\rho_{ee}^{(n=-1,0,1)},\rho_{gg}^{(n=-1,0,1)},\rho_{eg}^{(n=-1,0,1,2)}$, and $\rho_{ge}^{(n=-2,-1,0,1)}$. The analytical solution of the ground state population for various frequency terms are
\begin{widetext}
\begin{equation}\label{rdd}
\begin{aligned}
\rho_{dd}^{(-1)}=\frac{2i \Omega^2}{\frac{8 \Omega^2(\gamma-2i\omega_c)\left(\gamma^2+2\left(\delta_1^2+\delta_2^2\right)-4i\left(\gamma-i\omega_c\right)\omega_c\right)}{\left(\gamma -2 i \left(\delta _1+ \omega_c \right)\right)(\gamma -i \omega_c )  \left(\gamma +2 i \left(\delta _2-\omega_c \right)\right)}-\frac{3i \omega_c  \left(\gamma-2i \delta _1\right)\left(\gamma+2i\delta _2\right)}{2 \gamma-3i\omega_c}},\quad\rho_{dd}^{(0)}=\frac{1}{2},\quad\rho_{dd}^{(+1)}=\rho_{dd}^{(-1)*},
\end{aligned}
\end{equation}
\end{widetext}
and the total population in the ground states $\rho_{dd}=\rho_{dd}^{(0)}+(Re[\rho_{dd}^{(-1)}]+Re[\rho_{dd}^{(+1)}])\cos(\omega_c t-\phi_c)+(Im[\rho_{dd}^{(-1)}]-Im[\rho_{dd}^{(+1)}])\sin(\omega_c t-\phi_c)$ and $\rho_{cc}=1-\rho_{dd}$. The above ground state populations contain all pertinent aspects of the atomic response as the excited states populations are negligible for weak excitation. The subnatural feature of PO is significant only when $\omega_c\ll\gamma$, as seen in Fig.~\ref{rhocc2}(b). Thus, for small $\omega_c$, ie., $\delta=\delta_1\approx\delta_2$, we obtain from Eqn.~\eqref{rdd}
\begin{equation}
\begin{aligned}
\rho_{dd}^{(-1)}\approx\frac{4i\gamma\Omega^2}{16\gamma\Omega^2-3i\omega_c\left(\gamma^2+4\delta^2\right)},
\end{aligned}
\end{equation}
and the resulting Lorentzian has a full width at half maximum 
\begin{equation}\label{fwhm1}
\begin{aligned}
\Delta_{\text{FWHM}}=\frac{32\gamma\Omega^2}{3(\gamma^2+4\delta^2)}.
\end{aligned}
\end{equation}

Clearly, if OP governs the PO then the subnatural width $\Delta_{\text{FWHM}}$ (Eqn.~\eqref{fwhm1}) must be established as arising purely from OP. In order to show this, we obtain the OP rate independently by considering a single component (say $\hat{\sigma}_+$) of electric field ${\bm E}=E_0/(2\sqrt{2})\,\hat{\sigma}_+e^{-i\omega t}+c.c$ coupling the system. In steady state, the temporal dynamics of the density matrix elements will be of the form $\rho_{aa}(t)=\rho_{aa}e^{-\alpha t}$, $\rho_{dd}(t)=\rho_{dd}e^{-\alpha t}$, $\rho_{ad}(t)=\rho_{ad}e^{-\alpha t-i\omega t}$, and $\rho_{da}(t)=\rho_{da}e^{-\alpha t+i\omega t}$, where, $\alpha$ would be considered as the rate at which OP transfers the population from the ground state $|d\rangle$ to $|c\rangle$. On solving the Master equation, the analytical solution obtained for OP rate under weak excitation is
\begin{equation}\label{OPrate1}
\begin{aligned}
\alpha=\frac{4\gamma_{ac}|\Omega|^2}{((\gamma_{ac}+\gamma_{ad})^2+4\delta^2)}=\frac{4\gamma|\Omega|^2}{3(\gamma^2+4\delta^2)},
\end{aligned}
\end{equation}
Here, $\Omega=|d_{ad}|E_0/(2\sqrt{2})$ is the Rabi frequency of the single field excitation considered $(\sigma_+)$. For bichromatic excitation, the amplitude variation of the $\hat{\sigma}_{\pm}$ components can be written as $E_0/\sqrt{2}\,\cos(\omega_ct/2+\phi/2\pm\pi/4)$, thus, rescaling the Rabi frequency resulting spectral width for PO is 
\begin{equation}\label{fwhm2}
\begin{aligned}
\Delta=\frac{32\gamma|\Omega|^2}{3(\gamma^2+4\delta^2)},
\end{aligned}
\end{equation}
which is identical to Eqn.~\eqref{fwhm1}. Thus, we conclude that it is indeed OP that causes subnatural PO with the width defined by the OP rate.

\begin{figure}[ht]
\includegraphics[clip,trim=.1cm .1cm .1cm .1cm,width=\linewidth]{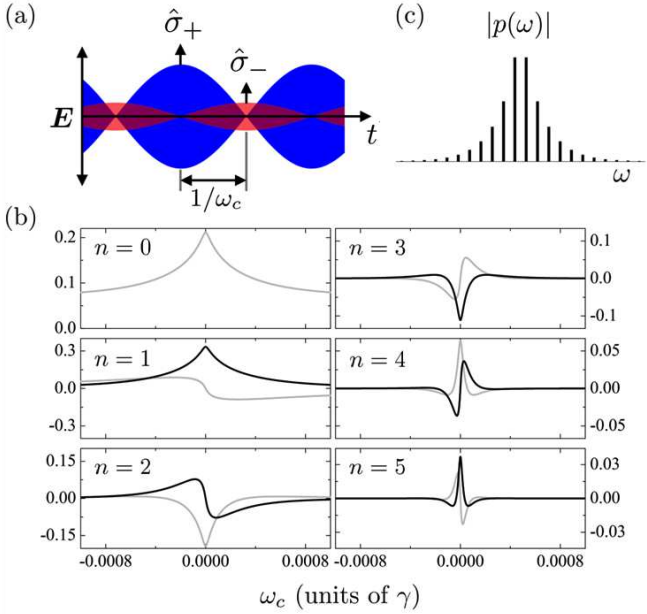}
\caption{(a) The $\hat{\sigma}_\pm$ components of elliptically polarized fields described in Eqn.~\eqref{ellipeq} for $\epsilon=30^{\circ}$. (b) Amplitude of various harmonics the in-phase (gray) and in quadrature (black) components of ground state population $\rho_{dd}$ for $\epsilon=30^{\circ}$ and $|\Omega|=.01\gamma$, (c) the magnitude of multiple sidebands contributing to the induced dipole moments in the adiabatic limit.}\label{ellip}
\end{figure}

The closed-loop phase $\phi_c$ becomes irrelevant for $\omega_1\neq\omega_2$, as it introduces at most an additional phase to the modulated field as well as to the generated frequencies. However, at $\omega_1=\omega_2$, the closed-loop phase plays a central role as the interference of the degenerate wave mixing processes governs the overall response. The interference of the induced dipole moments ${\bm p}(\omega_1)$ and ${\bm p}(\omega_1+2\omega_c)$ as well as ${\bm p}(\omega_2)$ and ${\bm p}(\omega_2-2\omega_c)$ depends directly on $\phi_c$, wherein, they may interfere constructively ($\phi_c=0$) and make the system akin to a two-level atom or destructively ($\phi_c=\pm\pi/2$) and make the medium completely transparent. When the initial phase difference between the two fields are chosen to be $\pi/2$ ($-\pi/2$), the destructive interference completely eliminates $\hat{\sigma}_-$ ($\hat{\sigma}_+$) component. As a result, all the population gets transferred and trapped in the bare state $|c\rangle$ ($|d\rangle$).

\begin{figure}[ht]
\includegraphics[clip,trim=.1cm .1cm .1cm .1cm,width=\linewidth]{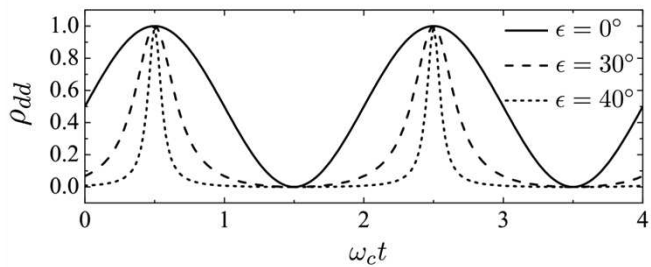}
\caption{Ground state PO for various input ellipticity $\epsilon$ in the adiabatic limit for weak excitation.}\label{ellip30}
\end{figure}

Furthermore, we can achieve temporal shaping of the PO by engineering the polarization content of the two fields that can lead to creation of multiple sidebands, while continuing to be at low light levels. Here, we consider two elliptically polarized fields ${\bm E}_1=E_0(\cos(\epsilon)\cos(k_1z-\omega_1 t)\hat{x}-\sin(\epsilon)\sin(k_1z-\omega_1 t)\hat{y})$ and ${\bm E}_2=E_0(\sin(\epsilon)\sin(k_2z-\omega_2 t)\hat{x}+\cos(\epsilon)\cos(k_2z-\omega_2 t)\hat{y})$ with the same ellipticity $|\tan(\epsilon)|$ and sense of rotation, however, their major axes are orthogonal. This can be recast in the atomic spherical basis as
\begin{equation}\label{ellipeq}
\begin{aligned}
{\bm E}_1&=\left(-\sin\left(\theta\right)\hat{\sigma}_++\cos\left(\theta\right)\hat{\sigma}_-\right)\frac{E_0}{2}e^{i(k_1z-\omega_1t)}+c.c\\
{\bm E}_2&=\left(-i\sin\left(\theta\right)\hat{\sigma}_+-i\cos\left(\theta\right)\hat{\sigma}_-\right)\frac{E_0}{2}e^{i(k_2z-\omega_2t)}+c.c.
\end{aligned}
\end{equation}
where, $\theta=\epsilon+(\pi/4)$. The  amplitude modulated $\hat{\sigma}_{\pm}$ components are shown in Fig.~\ref{ellip}(a), and their relative amplitude depends on  $\tan(\theta)$. We describe the dynamics of the atom using following nonlinear rate equations
\begin{equation}\label{rateeq}
\begin{aligned}
\dot{\rho}_{cc}=\alpha_+{\rho}_{dd}-\alpha_-{\rho}_{cc},\quad\dot{\rho}_{dd}=\alpha_-{\rho}_{cc}-\alpha_+{\rho}_{dd},
\end{aligned}
\end{equation}
where, $\alpha_{\pm}$ are the OP rates at which $\sigma_{\pm}$ components transfer the population between $\rho_{dd} \xrightleftharpoons[\sigma_-]{\sigma_+} \rho_{cc}$ as described earlier in Eqn.~\eqref{OPrate1}. For the input field considered in Eqn.~\eqref{ellipeq}, the OP rate given in Eqn.~\eqref{OPrate1} can be altered for the modified Rabi frequency as
\begin{equation}\label{OPrate}
\begin{aligned}
\alpha_{\pm}=\frac{8\gamma|\Omega|^2(1\mp\cos(2\theta))(1\mp\sin(\omega_ct))}{3(\gamma^2+4\delta^2)}.
\end{aligned}
\end{equation}
The closed-loop expansion (Eqn.~\eqref{WM1}) yields a recurrence relation between various harmonics of population
\begin{eqnarray}\label{recurrence}
&&i\cos(2\theta)\,\rho_{dd}^{(n-1)}-x_n\,\rho_{dd}^{(n)}-i\cos(2\theta)\,\rho_{dd}^{(n+1)}=\nonumber\\
&&\qquad\qquad\qquad\quad\cos^2(\theta)(i\delta_{n,1}-2\delta_{n,0}-i\delta_{n,-1}),\\
&&\hspace{-2em}\text{where, }x_n=2+\frac{in\omega_c(\gamma^2+4\delta^2)}{8\gamma|\Omega|^2}.\nonumber
\end{eqnarray}
We solve this recurrence relation with the condition $\rho_{dd}^{(n\to\pm\infty)}=0$. The steady state solution for $\rho_{dd}^{(0)}$ can be expressed as a continued fraction  
\begin{equation}\label{zero}
\begin{aligned}
\rho_{dd}^{(0)}=\cos^2(\theta)\frac{-\frac{\cos(2\theta)}{y_{-1}}+x_0-\frac{\cos(2\theta)}{y_{+1}}}{-\frac{\cos^2(2\theta)}{y_{-1}}+x_0-\frac{\cos^2(2\theta)}{y_{+1}}},
\end{aligned}
\end{equation}
and the complex amplitude of all other harmonics can be found using the recurrence relation
\begin{equation}\label{recurrence1}
\begin{aligned}
y_{\pm p}\,\rho_{dd}^{(\pm p)}=\mp i\cos^2(\theta)\,\delta_{p,1}\pm i\cos(2\theta)\,\rho_{dd}^{(\pm (p-1))},
\end{aligned}
\end{equation}
where, $p$ is a positive integer and
\begin{equation}\label{recurrence}
\begin{aligned}
y_{\pm p}=x_{\pm p}-\frac{\cos^2(2\theta)}{x_{\pm (p+1)}-\frac{\cos^2(2\theta)}{x_{\pm (p+2)}-\frac{\cos^2(2\theta)}{x_{\pm (p+3)}\cdots}}}.
\end{aligned}
\end{equation}
 The complex amplitude of individual harmonics of population as described by above equations are shown in Fig.~\ref{ellip}(b) for $\epsilon=30^{\circ}$ as $\omega_c$ varied. The number of harmonic contributions critically depends on the parameter $\cos(2\theta)$, and leads to generation of multiple sideband frequencies within a narrow band of frequency as shown in Fig.~\ref{ellip}(c). The interference of multiple phase locked harmonics results in the sharp temporal peaks in the PO as ellipticity increased and is shown in Fig.~\ref{ellip30}. This rate equation model that ignores the excited states matches exactly with the full Master equation treatment (Eqn.~\eqref{master}) for weak excitation, ie., $|\Omega|\ll\gamma$, but for arbitrary polarization content of the fields.

In conclusion, we have shown robust subnatural response in absence of QI and arising solely from OP, and the subnatural linewidth is fundamentally limited by the OP rate, and it can be further reduced by decreasing the intensity or by moving away from resonance. The rate equation model (Eqn.~\eqref{rateeq}) effectively captures the dynamics due to the lack of coherence or QI in the system. Definite control over the polarization content of the input bichromatic field allows temporal shaping of PO accompanied with nonlinear generation of multiple sidebands at low light levels. The large optical depth arising from \textit{complete} transfer of PO between the two ground states at low light levels holds immense promise in the control of occupation of $N$-atom Dicke states~\cite{Dicke}. The read-write protocol in such systems would be more realistic in photonic circuit applications as it does not involve saturating fields and offers manipulation through the polarization content of the fields~\cite{cpolambda4}.


\begin{thebibliography}{99}
\bibitem{ultranarrow1}
P. Zhou, and S. Swain, Phys. Rev. Lett. 77, 3995 (1996).
\bibitem{EIT}
M. Fleischhauer, A. Imamoglu, and J.P. Marangos, Rev. Mod. Phys. 77 633 (2005).
\bibitem{gsc1}
M.M. Kash, et al, Phys. Rev. Lett. 82, 5229 (1999).
\bibitem{gsc2}
D. Budker, D.F. Kimball, S.M. Rochester, and V.V. Yashchuk, Phys. Rev. Lett. 83, 1767 (1999).
\bibitem{CPO1}
S.E. Schwarz and T.Y. Tan, Appl. Phys. Lett. 10, 4-7 (1967).
\bibitem{CPO2}
M. Sargent, Phys. Rep. 43, 223-265, (1978).
\bibitem{CPO3}
L.W. Hillman,  R.W. Boyd, J. Krasinski, C.R. Stroud, Opt. Commun. 45, 416-419 (1983).
\bibitem{CPO4}
G.S. Agarwal, and T.N. Dey, Laser Photon. Rev. 3, 287-300 (2009).
\bibitem{cpolambda1}
T. Laupr{\^e}tre, et al, Phys. Rev. A 85, 051805 (2012).
\bibitem{cpolambda2}
A.J.F de Almeida, et al, Phys. Rev. A 90, 043803 (2014).
\bibitem{cpolambda3}
M.A. Maynard, F. Bretenaker, and F. Goldfarb, Phys. Rev. A 90, 061801 (2014).
\bibitem{cpolambda4}
P. Neveu, et al, Phys. Rev. Lett. 118, 073605 (2017).
\bibitem{opticalp1}
W. Happer, Rev. Mod. Phys. 44, 169 (1972).
\bibitem{opticalp2}
R. Bernheim, ``Optical Pumping An Introduction" (Benjamin, 1965)
\bibitem{kaniEPL}
A kani and H. Wanare, Manuscript under consideration in EPL.
\bibitem{Dicke}
I.E Linington, and N.V. Vitanov, Phys. Rev. A 77, 010302 (2008).
\end{thebibliography}
\end{document}